# Effects of the baroclinic adjustment on the tropopause in the NCEP-NCAR reanalysis

Alessandro Dell'Aquila [1], Paolo M. Ruti[1] and Alfonso Sutera[2].

[1] *ENEA C.R. Casaccia, Rome, Italy*

[2]*Department of Physics, University of Rome "La Sapienza", Rome, Italy*

Corresponding author:

Dr. Alessandro Dell'Aquila

ENEA, C.R. Casaccia,

Via Anguillarese 301, 00060 S. Maria di Galeria, Rome, Italy

E-mail: alessandro.dellaquila@casaccia.enea.it

Phone:+39-06-30481072

Fax:+39-06-30484264

# Abstract

In this work, we study the mean tropopause structure from the NCEP-NCAR reanalysis in the framework of a theory of baroclinic adjustment, focusing on the impact of baroclinic eddies on the mean tropopause height.

In order to measure the effects of such perturbations, we introduce an appropriate global index that selects events of high baroclinic activity and allows us to individuate the phases of growth and decay of baroclinic waves. We then composite the tropopause mean structure before and after a baroclinic event, finding that baroclinic disturbances cause the zonally averaged midlatitude winter tropopause height to rise. Our results suggest that the baroclinic adjustment processes should be taken into account to explain the nature of the midlatitude tropopause.

# 1.Introduction

The mechanisms which concur to maintain and modify the tropopause structure constitute a relevant issue in the investigation and understanding of the climate system. A classical view of climate change induced by external factors (Manabe and Wetherald, 1967) relates planet surface temperature modifications to changes of the tropopause height. In this view, both radiative-convective balance in the troposphere and the radiative equilibrium in the lower stratosphere concur to determine the vertical structure of the temperature, and, consequently, the tropopause height. Thus, for fixed stratospheric temperature profile and tropospheric lapse rate, a rise in the mean tropopause corresponds to a warmer mean surface temperature.

Recent studies have built on this approach (Hoinka, 1998; Santer et al., 2003). In particular, Santer et al. (2003) have suggested that, besides surface temperature, tropopause pressure may be a good fingerprint of climate change. Some negative trends on global and regional means of the tropopause pressure (corresponding to an increase in the tropopause height) have been detected in the available reanalysis (NCEP-NCAR), especially in the last two decades. These signals could be due to human-induced changes in ozone and well-mixed greenhouse gases, as simulated in GCMs runs (Santer et al., 2003), where a trend in the global tropopause pressure is recognized only if these anthropogenic modifications are considered. Therefore, an analysis of the tropopause structure and of the processes which concur to determine it will contribute both to a more accurate comprehension of the climate system, and to the evaluation of the effects induced by an external stress on the climatic equilibrium.

The simplest model capable of explaining the existence of the tropopause is one based on a radiative–convective balance (e.g., Houghton, 1977), particularly plausible for the tropical tropopause. The radiative equilibrium temperature near the ground is adjusted by convective activity to a statistically neutral state, while the profile in higher altitudes is statistically stable. Out of the tropical region, the physical mechanisms determining the tropopause maintenance and location are still object of investigation. However, it is recognised in the recent literature (see, among others, Egger, 1995; Thuburn and Craig, 1997; Haynes et al, 2001) that the structure of the extra-tropical tropopause may be influenced by baroclinic adjustment processes. Green (1970) and Stone (1978) were the first to propose an adjustment mechanism, referred to as "baroclinic adjustment", which maintains the mean atmospheric state close to neutral stability when large-scale forcing tends to make the mean state baroclinically unstable. In the last decades, several theories (see, e.g., , Lindzen and Farrell, 1980; Held, 1982; Lindzen, 1993; Bordi et al., 2002, 2004; Dell'Aquila 2004) have been proposed concerning the mechanisms that limit the growth of the baroclinically unstable eddies, modify the basic state where the unstable modes grow, and bring the system to equilibrium. These theories can be viewed as a parameterisation of the effects of baroclinic eddies on the general circulation and on the mean atmospheric structure in the midlatitudes.

Baroclinic perturbations transport heat meridionally and vertically, modifying the mean vertical structure of the extratropical troposphere and, hence, the height of the extratropical tropopause. Thus, the structure of the baroclinically readjusted state may be thought to provide a constraint on the mean tropopause height. This was made explicit by Held (1982), who introduced a "dynamical constraint" relating the tropopause height and the surface temperature, based on the Charney and Stern (1962) baroclinic stability condition for a symmetric basic state. A different description of the baroclinic neutral state, based on a geometric adjustment of the tropopause height, was proposed by Lindzen (1993). In this theory the height of the tropopause neutralising all unstable modes is defined as a function of the jet-width, the Brunt-Vaisala frequency and the Coriolis parameter. Lindzen's theory has recently been extended by Bordi et al. (2002; 2004) to include the effects of the stratosphere and of the orography.

The main purpose of the present paper is to supply observational evidence of the importance of the baroclinic adjustment in determining the tropopause structure. In particular, we shall evaluate the impact of intense baroclinic activity on the mean tropopause pressure from global reanalysis, by comparing the midlatitude tropopause fields before and after baroclinic events.

The work is structured as follows. Section 2 describes the data used in this study, while in Section 3 we construct an index which measures the intensity of the baroclinic activity at 500 hPa, and allows us to individuate periods of growth and decay of the baroclinic perturbations. Section 4 analyses differences in the winter tropopause height before and after baroclinic events. Finally, a review of the main results is given in the Summary.

# 2. Data

For this study, we use the data freely provided by the NCEP-NCAR Archive. In order to provide a constant climate monitoring, NCEP (National Center for Environmental Prediction) and NCAR (National Center for Atmospheric Research) are cooperating to produce a record of global reanalyses of atmospheric fields, performed using a global spectral model with T62 horizontal resolution and 28 vertical sigma levels (Kalnay et al, 1996), in which data from land surface, ship, rawinsonde, aircraft and satellite are assimilated. The system control for the quality and the assimilation of data is kept unchanged over the reanalysis period from 1948. However, in the earliest decade (1948-1957) upper air data observations were fewer and primarily in the Northern Hemisphere, so that the reanalysis for this period is less reliable than for the later 40 years (Kistler et al. 2001). For our purposes we use the daily data on a 2.5° X 2.5° latitude-longitude grid for the northern winter period (DJF) from 1st December 1957 to 28th February 2001.

To construct an indicator of baroclinic activity and to verify if it might identify baroclinic events, we use the daily 500 hPa geopotential height and air temperature, and the 700hPa zonally averaged meridional heat flux. Concerning the tropopause field, we use the daily tropopause pressure provided by NCEP-NCAR, computed from the vertical temperature profile. The tropopause is defined as the lowest level where the temperature lapse rate becomes less than 2K/km, as defined by the WMO, 1957 (thermal tropopause).

The thermal tropopause pressure from the NCEP-NCAR archives is allowed to vary between 450hPa and 85 hPa. In the classification described in Kalnay et al.

(1996), the tropopause pressure is placed in the "A" (first) class of reliability regarding the reanalysis gridded fields. As observed by Kalnay et al. (1996), this kind of data are strongly influenced by the observations and weakly affected by the model used for the assimilation.

# 3. Constructing a global index of baroclinic activity

The baroclinic waves transport heat and momentum vertically and horizontally at synoptic scales in the midlatitudes, gaining available potential energy from the zonally symmetric flow. These disturbances explain most of the high frequency variability in the extratropics. They are typically eastward propagating and characterized by a period shorter than 10 days and a zonal wave number $6 \leq k \leq 8$ (Fraedrich and Bottger, 1978).

To identify the most active baroclinic disturbances, we use a global indicator similar to the Wave Amplitude Index (WAI) introduced by Sutera (1986) or the baroclinic index by Benzi and Speranza (1989). .

Here follows a short description of the steps to construct our index:

- firstly, we apply a high-pass filter to the geopotential daily field, in order to retain only the modes with a period shorter than 10 days;
- we then average the 500 hPa geopotential height field $z(\lambda, \phi)$ over the latitudinal band bounded by 30°N and 60°N;
- for each day in the DJF period, 500 hPa geopotential height is Fourier decomposed in the longitudinal direction $\lambda$;
- the index is finally computed from the variance associated to the Fourier coefficients $Z_k$ of the zonal wavenumbers $k$=6-8:

$$Z_{6-8}(t) = \left( \sum_{k=6}^{8} 2|Z_k(t)|^2 \right)^{\frac{1}{2}}. \tag{1}$$

Some remarks are in order. First of all, the index computed as a function of latitude, $Z_{6-8}(t, \phi)$, shows a maximum of variance at about 45°N, as shown in

Fig.1. Thus, the choice of the latitude band 30°N-60°N allows us to capture the bulk of the baroclinic activity. We also note that the probability distribution function of the index (here not shown) is essentially unimodal. Secondly, the use of the low order Fourier zonal components corresponds to focusing on global baroclinic instabilities, whose relevance has been recently investigated by several authors (see, among others, Pierrehumbert,1984; Lee and Mak, 1995).

By using our index we can select the periods of intense baroclinic activity. Operationally, we fix a LOW and a HIGH threshold, $Z_L$ and $Z_H$, respectively. We set:

- $Z_L = \overline{Z_{6-8}} - \frac{1}{2}\sigma_{Z_{6-8}}$ ; (2)

- $Z_H = \overline{Z_{6-8}} + \frac{1}{2}\sigma_{Z_{6-8}}$ , (3)

where the over bar denotes the time mean and $\sigma$ the standard deviation. When the $Z_{6-8} < (>) Z_L (Z_H)$ baroclinic activity is considered to be weak (strong). We state the troposphere undergoes a baroclinic event if $Z_{6-8}$ becomes higher than $Z_H$ at least for two days after having been lower than $Z_L$ for at least two days. We define this as the IN phase. Similarly, the atmosphere leaves the baroclinic enhanced state if $Z_{6-8}$ becomes lower than $Z_L$ at least for two days after having been higher than $Z_H$ for at least two days. We define this as the OUT phase. For our choice of the threshold values, we detected approximately 100 events in the 44 northern winters we analysed.

The patterns of the disturbances in the growing and decaying phases of the baroclinic events are shown in Fig.2a-f. In particular, we show the composites of the 500 hPa geopotential height and air temperature. We composite the beginning days of baroclinic event (IN), the days before (IN-1), and the days after (IN+1). The same has been done for the ending phases of the perturbations (days OUT, OUT-1,OUT+1). We applied to both fields a time high-pass (days<10) filter and a spatial filter that retains only the zonal wave numbers *k=6-8*. On days IN-1,IN, IN+1, it is evident from Fig. 2a-c that a baroclinic event is growing in the Pacific and Atlantic storm tracks regions. The disturbances in the geopotential height and air temperature fields, that are not in phase, increase in amplitude and propagate eastward. When the baroclinic event is decaying, on days OUT-1, OUT, OUT+1, the perturbation is located over the Atlantic region, where the disturbance dies out (Fig. 2d-f). On the OUT days the phase shift between the two fields is less evident, causing the baroclinic conversion to be less efficient and the net poleward heat flux to decrease. On the OUT+1 days the baroclinic event has ceased.

We can also argue from the composites of Fig.2 that this kind of perturbations has a non-zero component with the spatial phase locked by the large scale orography.

In order to check the reliability of our index we perform a further analysis. The main feature of a baroclinic disturbance is the poleward and vertical heat transport. Thus, the baroclinic nature of the selected events may be corroborated by an analysis of these quantities. The zonally averaged poleward heat flux is obtained by averaging the term $v_{6-8} T_{6-8}$ over the latitudinal belt 30°N-60°N and then on the longitudinal direction. Here, $v_{6-8}$ and $T_{6-8}$ are the meridional wind velocity and air temperature at 700 hPa, and we apply a time high-pass filter and a spatial filter as for the 500 hPa fields of Fig. 2. We choose the vertical level where the baroclinic meridional heat transport is maximum, and the latitudinal band representative of the mid-latitude dynamics. The resulting zonally averaged poleward heat fluxes are composed for the events we isolated using the $Z_{6-8}$ index. In Fig. 3 we present the probability density function (PDF) for the amplitude of the poleward heat flux for the seasonal average DJF (thick line), the IN+1 days (thin line) and the OUT+1 days (dashed line). When the baroclinic event occurs (IN+1 days) the PDF of the heat flux shows a well defined maximum at an amplitude higher than in the other two cases. These results show the reliability of the $Z_{6-8}$ index as a proxy of global baroclinic activity.

# 4. Effects on the mean tropopause structure

To evaluate the effects of baroclinic waves on the tropopause we analyze the tropopause profiles before and after the events selected by the procedure introduced in the previous Section. A reasonable hypothesis is that these disturbances leave the atmosphere close to a state of baroclinic neutrality (Bordi et al. 2002; 2004). Thus, the tropopause observed after a baroclinic event (OUT phase) should be close to the height corresponding to the baroclinic neutral state. The tropopause difference between the baroclinic neutral state (after the event) and the unstable state (before the event) should provide an estimate of the effects of the baroclinic adjustment on the tropopause.

In Fig. 4, we show the zonally averaged midlatitude tropopause pressure for the IN and OUT phases as a function of latitude (the DJF and JJA meridional profiles are also reported for comparison). A detailed description of climatological meridional profiles of the tropopause is provided by Hoinka (1998) and Dell'Aquila (2004) . We focalise on the tropopause profiles in the midlatitude regions, where the baroclinic waves are active. We compare the tropopause pressure averaged over the IN phase (thin solid line) with the one averaged over the OUT phase (dashed line). At lower latitudes, ($\phi<45°N$), the IN and OUT profiles are mostly coincident with the seasonal profile (thick solid line). In this region baroclinic waves do not have a mean effect on the tropopause meridional profile, as well as in the polar region ($\phi>70°N$). Instead, in the midlatitude band ($45°N\leq\phi\leq65°N$) the observed tropopause is deeply affected by baroclinic perturbations. As clearly shown in the figure, before the baroclinic event, the tropopause is generally lower with respect both to the readjusted state of the OUT days and to the mean seasonal DJF profile. In order to estimate the statistical significance of these results, we apply a *t*-test with 95% confidence level to the meridional tropopause profiles reported in Fig.4. The null hypothesis is that these states have the same distribution. In the dotted box ($47.5°N \leq \phi \leq 67.5°N$) the null hypothesis could be rejected and the differences obtained are proved to be significant. The utmost difference in tropopause pressure between IN and OUT days is about 8 hPa which roughly corresponds to a variation of about 400 meters in midlatitude standard atmosphere. The difference between the DJF and the JJA profiles is about 30-35 hPa in midlatitudes and it measures the radiative constraint due to the seasonal cycle, but also takes into account the effect of the wintertime baroclinic adjustment itself. So, the 8 hPa difference shown in Fig. 4 seems to be relevant.

The latitude where the effect of the disturbances on the tropopause is more profound does not correspond to the region where the maximum of the baroclinic activity occurs (around 45°N, as shown in Fig. 1). In fact, the differences between the tropopause pressure in the IN and the OUT phases are more evident in the northern part of the midlatitude channel ($55°N \leq \phi \leq 65°N$). A theoretical explanation of this effect in the context of a simple model has been advanced by Egger (1995). He argued that the vertical and meridional heat transports associated to baroclinic unstable waves, simply depicted as Eady waves, would cause a rise of the zonal mean tropopause height in the northern part of a $\beta$-channel, by warming the tropospheric layers below the tropopause and cooling the lowermost stratosphere.

The later feature is confirmed by the analysis of the time autocorrelation function performed on times series of the $Z_{6-8}$ index and of the zonally averaged tropopause pressure computed over three different latitude bands: 40°-45°N ; 55°N-65°N ; 70°N-75°N. Moreover, we repeat the analysis on the region of the maximum baroclinic effect (55°N-65°N) considering only the days after baroclinic events. The corresponding results are shown in Fig.5. We can observe that, near the tropical and polar regions (thin solid and dashed lines, respectively), the zonal mean tropopause pressure is driven by a forcing with a characteristic time scale of about 20 days or more. Over the tropical band the atmosphere tends towards a radiative-convective equilibrium, while over the polar region the radiative constraint dominates. In the region 55°N-65°N (thick solid line) a faster decay takes place; the autocorrelation function decreases very quickly over the first few days. Furthermore, if we just consider the readjustment after baroclinic events in this band, the autocorrelation functions of the zonal mean tropopause (dot-dashed thick line) and of the $Z_{6-8}$ index (dot-dashed line), both showing a very fast decay, closely resemble each other. The decay time scale of a few days is typical of baroclinic instability. Therefore, the zonal mean tropopause dynamics seems to be characterized by processes related to radiative-convective equilibrium and baroclinic adjustment that concur, at different time scales, to fix and maintain the observed mean state.

In order to analyse the zonal mean tropopause behaviour during a baroclinic life-cycle, we composite the anomaly of the zonally averaged tropopause pressure from day OUT-4 to day OUT+4 as a function of latitude (Fig. 6). The anomaly is computed with respect to the mean IN days. A clear rise of the tropopause appears in the northern part of the channel over the last days of the baroclinic event and then it moves poleward in the following days. When the disturbance ceases the readjusted tropopause stay higher for at least 4 more days. Then, the radiative forcing acts to lower the tropopause and the anomaly tends to vanish.

## 5. Summary

In this work, we have investigated the effects of wintertime baroclinic eddies on the mean tropopause pressure, in the NCEP-NCAR reanalysis. We have introduced a global index of baroclinic activity, that has proven to be a reliable indicator of global baroclinic events. Using this index we can identify periods in which baroclinic perturbations grow (IN phase) and decay (OUT phase). The comparison between the tropopause pressures over the IN and the OUT phases has allowed us to estimate the effect of baroclinic eddies.

We have found that baroclinic events cause a significant rise in the mean observed tropopause at midlatitudes, in agreement with some recent theoretical investigations of baroclinic adjustment in simple models ( Lindzen, 1993; Egger, 1995; Bordi et al., 2002, 2004; Dell'Aquila, 2004), proposing a tropopause readjustment to a level higher than that fixed by the radiative convective equilibrium. This effect is evident in the zonally averaged structure of the tropopause, in particular for the latitude band $55°N \leq \phi \leq 65°N$, where the time scale of the zonal mean tropopause is very close to the characteristic time scale of baroclinic instability. We also found that the tropopause adjustment is more evident in the northern part of the 'midlatitude channel', in agreement with theoretical predictions by Egger (1995).

These results show that taking into account the adjustment to a baroclinic neutral state allows to better describe the behaviour of the midlatitude mean tropopause, providing a "dynamical constraint" to add to the usual and well-known "radiative constraint" (Held, 1982; Thuburn and Craig, 2000).

A useful future step will be to verify if the effects of baroclinic disturbances we have observed in this work may be correctly reproduced in GCM simulations.

These effects may also in studies on the sensitivity of the mean tropopause structure to external (natural or anthropogenic) stresses, such as eventual changes in the stratospheric structure.

**Acknowledgements**

We would like to thank S. Calmanti, G. Pisacane, and M. Petitta for their helpful comments.

Data used in this study are NCEP-NCAR re-analysis data and have been provided by NOAA-CIRES Climate Diagnostics Center, Boulder, Colorado.

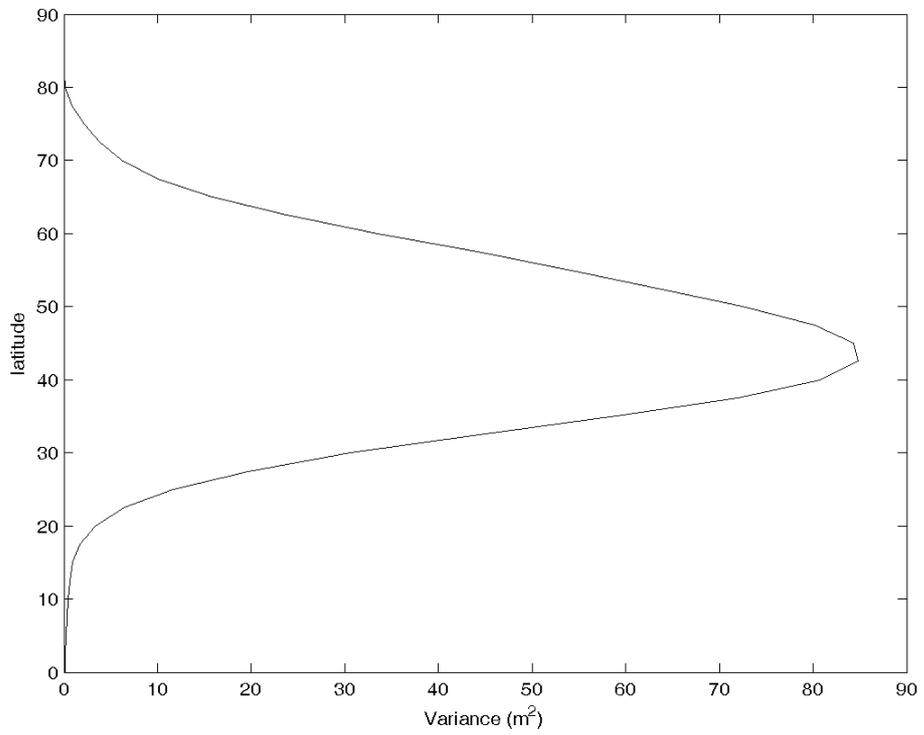

**Fig.1:** Variance of $Z_{6-8}$ as a function of latitude for DJF period in northern hemisphere

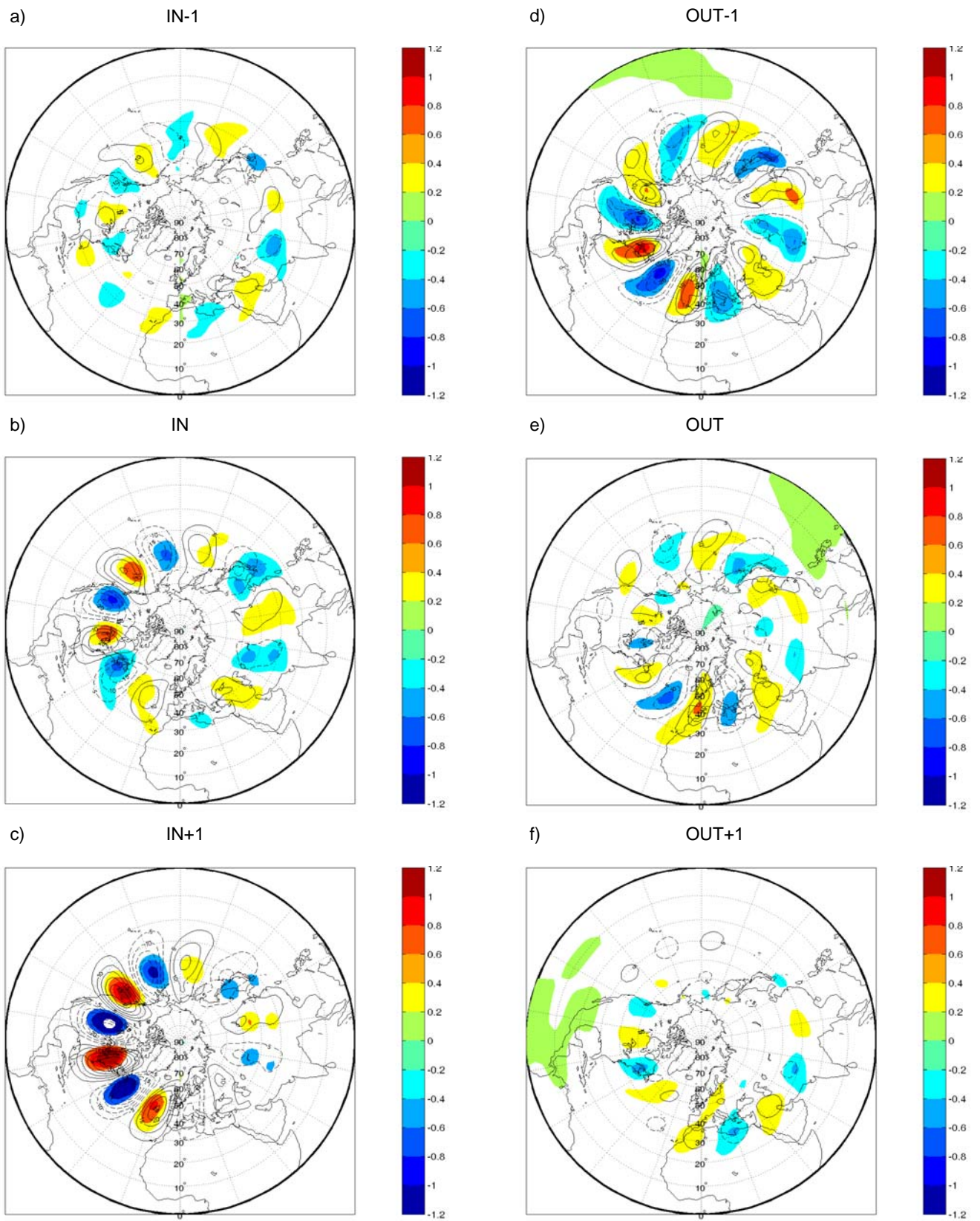

**Fig.2:** Composites of 500 hPa geopotential height in meters (contour lines) and 500 hPa temperature in Kelvin (fill contours) with a time band-pass and a zonal filter 6-8 in the DJF period for the days: IN-1 (a), IN (b), IN+1(c), OUT-1 (d),OUT(e),OUT+1 (f).

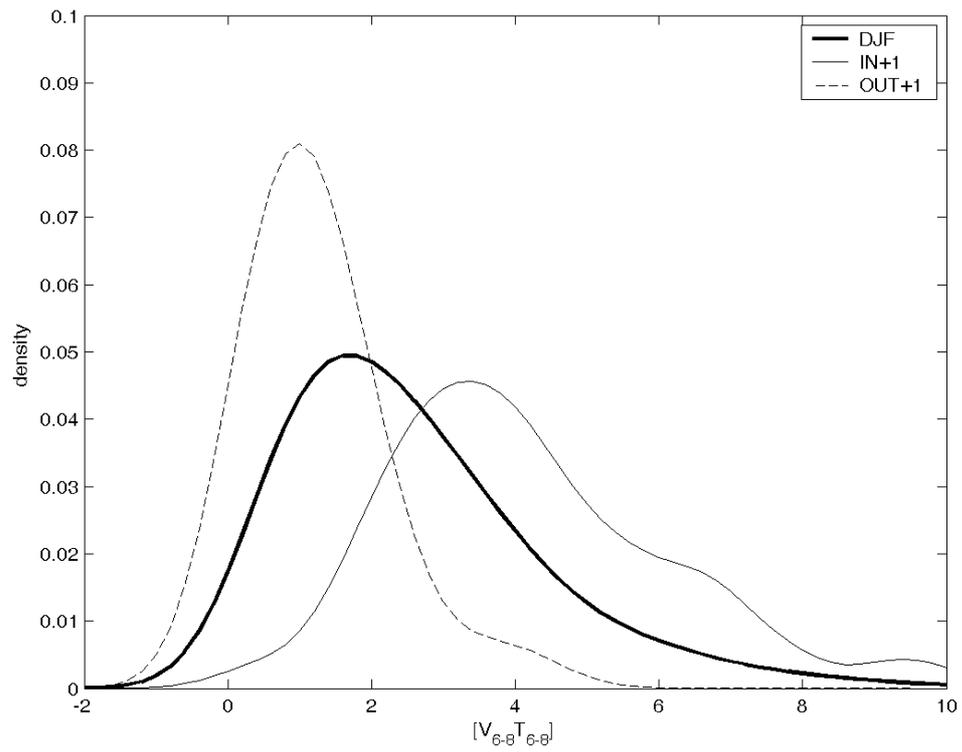

**Fig.3**: Probability density estimate of the amplitude of 700 hPa zonally averaged poleward heat flux [$v_{6-8}\, T_{6-8}$] in $K\, m\, s^{-1}$ for IN+1 and OUT+1 days. The seasonal average is also reported for comparison.

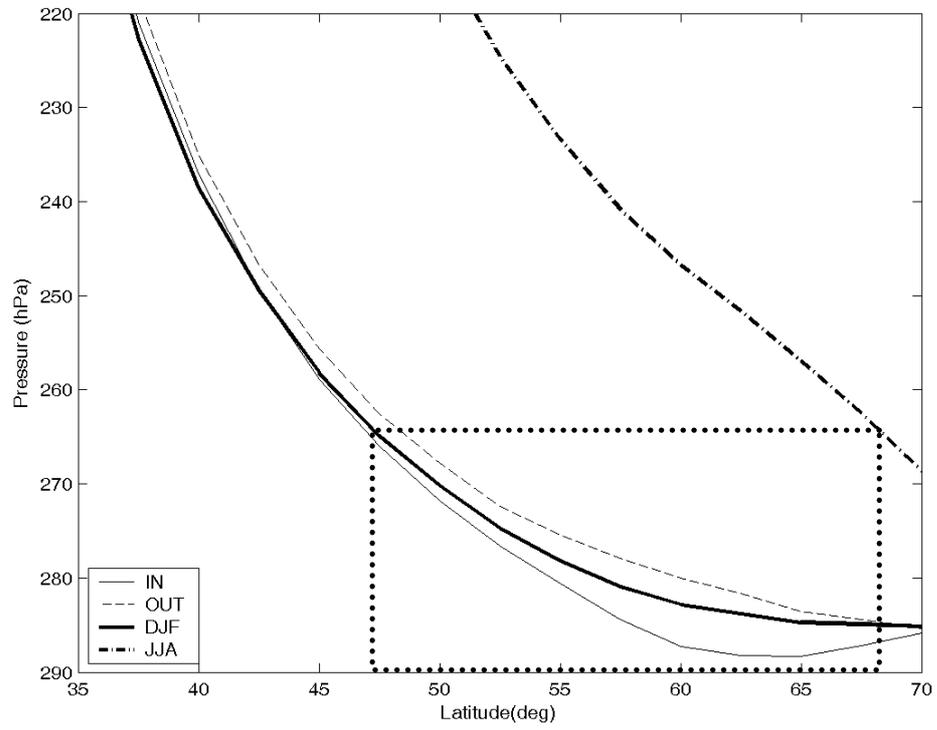

**Fig.4:** Meridional profiles of the zonally averaged tropopause pressure. In the dotted box the differences between IN and OUT profiles are significant with 95% confidence level (see the text for major details).

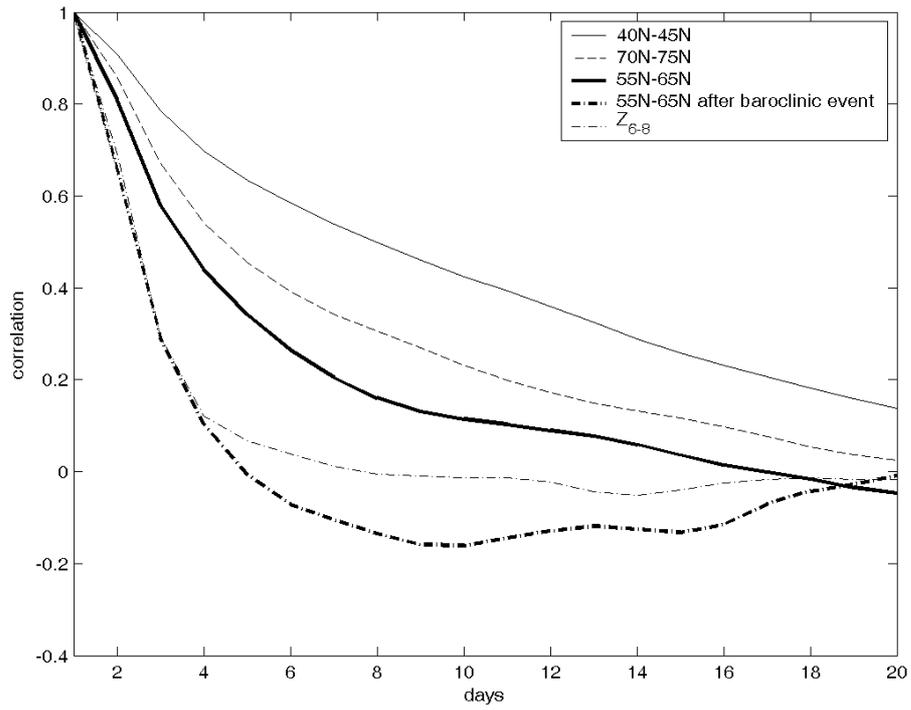

**Fig.5:** Time autocorrelation function for the zonal mean tropopause pressure averaged over three different latitude bands. We also report the autocorrelation function for the zonal mean tropopause pressure averaged over 55°N-65°N only after baroclinic events and for $Z_{6-8}$.

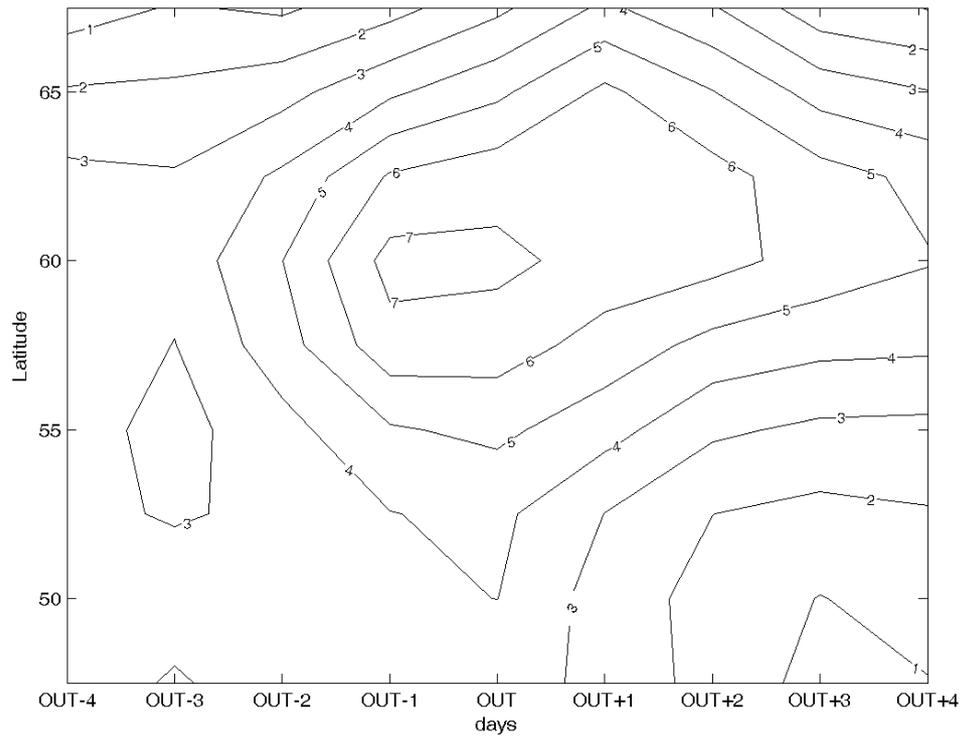

**Fig.6:** Anomaly in hPa (with respect to the IN days) of the zonally averaged tropopause pressure from OUT-4 to OUT+4 days as a function of latitude.